\documentclass{pasa}
\usepackage{graphicx}
\usepackage{amsmath}
\usepackage{gensymb} 
\usepackage{xfrac}
\usepackage[flushleft]{threeparttable}
\usepackage{amssymb}

\usepackage{aas-macros}
\usepackage{hyperref}
\hypersetup{colorlinks,citecolor=blue,linkcolor=blue,urlcolor=blue}
\usepackage{acronym}

\acrodef{AGN}{active galactic nuclei}
\acrodef{VLSSr}[VLSSr]{Very Large Array Low-frequency Sky Survey Redux}
\acrodef{LOTSS}[LoTSS]{LOFAR Two-metre Sky Survey}
\acrodef{MWA}{Murchison Widefield Array}
\acrodef{LOFAR}{LOw-Frequency ARray}
\acrodef{NVSS}{NRAO VLA Sky Survey}
\acrodef{SDSS}{Sloan Digital Sky Survey}
\acrodef{GLEAM-X}{GaLactic and Extragalactic All-sky Murchison Widefield Array eXtended survey}
\acrodef{GLEAM}{GaLactic and Extragalactic All-sky Murchison Widefield Array survey}
\acrodef{CAMB}[CAMB]{Code for Anisotropies in the Microwave Background}
\acrodef{ISW}[ISW]{Integrated Sachs-Wolfe effect}
\acrodef{CMB}{Cosmic Microwave Background}
\acrodef{RACS}[RACS]{Rapid ASKAP Continnum Survey}
\acrodef{FLASK}[FLASK]{Full-sky Lognormal Astro-fields Simulation Kit}
\acrodef{EoR}[EoR]{Epoch of Re-ionisation}
\acrodef{COBE}[COBE]{COsmic Background Explorer}
\acrodef{TGSS}[TGSS]{TIFR GMRT Sky Survey}
\acrodef{WMAP}[WMAP]{Wilkinson Microwave Anisotropy Probe }
\acrodef{LoTSS}[LoTSS]{LOFAR Two-metre Sky Survey}
\acrodef{SKADS}[SKADS]{Tiered Radio Extragalactic Continuum Simulation}
\acrodef{ACF}[ACF]{angular correlation function}
\acrodef{GLASS}[GLASS]{Generator for large-scale Structure}
\acrodef{SKA}[SKA]{Square Kilometre Array}
\acrodef{CAMB}[CAMB]{Code for Anisotropies in the Microwave Background}
\acrodef{SUMSS}[SUMSS]{Sydney University Molonglo Sky Survey}
\acrodef{APS}[APS]{angular power spectrum}
\acrodef{GXDS}[GXDS]{\ac{GLEAM-X} Data Subset}
\acrodef{LoFAR}[LOFAR]{Low Frequency Array}
\acrodef{GMRT}[GMRT]{Giant Meterwave Radio Telescope}
\acrodef{SKADS}[SKADS]{European SKA Design Study Simulated Skies}

\title[The Angular Correlation Function of GLEAM-X]{The Angular Correlation Function as measured by the GLEAM-X Survey}

\author[Venville et al.]
{Brandon Venville$^{1}$\thanks{email:brandon.venville@postgrad.curtin.edu.au}, 
David Parkinson$^{2}$, 
Natasha Hurley-Walker$^{1}$, 
Tim Galvin$^{3}$, 
Kathryn Ross$^{1}$
\affil{$^{1}$ICRAR-Curtin, Curtin University, Bentley, 6102, Western Australia, Australia}%
\affil{$^{2}$Korea Astronomy and Space Science Institute,  Daejeon 34055, Republic of Korea}%
\affil{$^{3}$ATNF, CSIRO Space \& Astronomy, Bentley, WA, Australia}%
}

\jid{PASA}
\doi{10.1017/pas.\the\year.xxx}
\jyear{\the\year}

\hypersetup{draft}

\begin{document}
\begin{frontmatter}
\maketitle
\begin{abstract}
    The angular correlation is a method for measuring the distribution of structure in the Universe, through the statistical properties of the angular distribution of galaxies on the sky. We measure the angular correlation of galaxies from  the second data release of the \ac{GLEAM-X} survey, a low-frequency radio survey covering declinations below $+30^\circ$. We find an angular distribution  consistent with the $\Lambda$CDM cosmological model assuming the best fitting cosmological parameters from \cite{Plank2020cosmo}. We fit a bias function to the discrete tracers of the underlying matter distribution, finding a bias that evolves with redshift in either a linear or exponential fashion to be a better fit to the data than a constant bias. We perform a covariance analysis to obtain an estimation of the properties of the errors, by analytic, jackknife and sample variance means. Our results are consistent with previous studies on the topic, and also the predictions of the $\Lambda$CDM cosmological model.
\end{abstract}

\end{frontmatter}

\section{Introduction}
        The large-scale structure evident in the matter distribution of the universe is predicted by the underlying cosmology. In the currently accepted $\Lambda$CDM model, the universe began with the initial singularity, following which there was a period of rapid inflation. In the primordial field driving this inflation, there were quantum fluctuations, seeding small density variations in the early universe. The evolution and properties of these perturbations are related to the cosmological parameters, as the evolution of these perturbations in the otherwise homogeneous early universe are governed by various mechanisms, such as the collapse due to gravity, the expansion of the universe, and the propagation of the density fluctuations through the primordial medium. As such, one can use measurements of the large-scale structure inherent in galaxy surveys to test consistency with the values of the cosmological parameters derived from other methodologies, such as observing the \ac{CMB}
        \citep{Planck2013Overview}.

        In recent years there has been renewed interest in radio astronomy, and in particular low frequency radio astronomy, as instruments such as \ac{LOFAR} and the \acl{MWA} \citep[MWA; ][]{Tingay2013,Wayth2018} perform large area sky surveys with unprecedented sensitivity. The \ac{MWA} is a low-frequency radio telescope located at Inyarrimanha Ilgari Bundara / the Murchison Radio-astronomy Observatory, operating at a frequency range of over 72--231\,MHz, and is the instrument used in this study.
        
        Prior studies have made successful measurements of the clustering inherent in the large-scale structure, through the \ac{ACF} as well as other means. This has included studies with radio surveys, such as the work of \cite{Dolfi2019} using \acl{TGSS} \citep[TGSS; ][]{TGSS}, the work of \cite{Blake2002} using the \acl{NVSS}\citep[NVSS;][]{1998AJ....115.1693C}, and more recently that of \cite{2024MNRAS.527.6540H} using \acl{LoTSS} \citep[LoTSS; ][]{LoTSS} and that of \cite{DavidISW} using \acl{RACS} \citep[RACS; ][]{2020PASA...37...48M,2021PASA...38...58H}. After accounting for survey and source specific factors such as sky coverage, resolution and the bias inherent in the observed tracers of the matter distribution, these studies have shown consistency with the predictions of the $\Lambda$CDM cosmological model. 

        In this study, the \ac{GLEAM-X}, is used to measure the clustering of radio tracers, primarily \ac{AGN}, through measuring the \ac{ACF}. In section~\ref{Theoretical considerations}, the theoretical basis of the \ac{ACF} is discussed. Following this, the methodology of this study shall be detailed in section~\ref{methodology}, with theoretical predictions discussed in section~\ref{theoretical estimation}. Covariance estimation is considered in section~\ref{covariance}. The results of this study and conclusion are summarised in sections~\ref{Resukts} and \ref{conclusion} respectively.
\section{Theory}\label{Theoretical considerations}
    \subsection{Power spectrum and cosmology}
    In the current $\Lambda$CDM model of cosmology, the large-scale structure of the universe evident today evolved from the collapse of density fluctuations seeded in the very early universe \citep{Peacock}. The theory of cosmological perturbations and their evolution is well developed \citep[see e.g.][]{1995ApJ...455....7M,1966ApJ...145..544H,1986ApJ...304...15B}. Essentially, the initial density perturbations are generated during the early inflationary epoch, before their growth is affected by gravitational collapse, the expansion of the universe, and radiation pressure from the collapsing matter. To derive a theoretical consideration incorporating these factors, the boltzmann code \ac{CAMB} \citep{CAMB2} was used, to evolve the perturbation equations to provide theoretical predictions for the angular correlation function for a given cosmology: this is further discussed in section \ref{theoretical estimation}.
    
    \subsection{Bias}\label{Bias}
    In using radio sources to trace large-scale structure one must be mindful of the bias of various source populations in tracing the underlying matter density. In this case, bias refers to the degree to which the tracer population follows the matter distribution, specifically:
    \begin{equation}\label{biass}
        \delta_g=b\delta_m \,.
    \end{equation}
    where $\delta_g$ is the perturbation in the galaxy density when compared to a homogeneous background, $\delta_m$ is the peturbation in the matter density, and $b$ is the bias of the tracer. Models for the bias of varying galaxy types have been proposed in the literature, whether derived from N-Body simulations \citep{Millenium2}, analytically \citep{Sheth,Paolo}, or empirically derived from observations \citep{DavidISW}. The most relevant of these in our case is that of \cite{2015ApJ...812...85N}, which was found by \cite{LoTSSW} to fit their \ac{ACF}, with \ac{LoTSS} also being a low frequency continuum survey.

    \subsection{\ac{ACF} and \ac{APS}}
    A radio survey such as \ac{GLEAM-X} cannot be used to measure the 3-dimensional power spectrum directly, as the redshifts of the sources are unknown. The 2-dimensional angular power spectrum is observable however, being the projection of the 3-dimensional spectrum onto the celestial sphere.
    The density perturbations of equation \ref{biass} are related to the 3 dimensional power spectrum by
    \begin{equation}
        P\left(k\right) = \langle \delta_k\rangle^2
    \end{equation}
    where $\delta_k$ is the ensemble average power of wavemode k.
    The theoretical \ac{APS} was derived, following \citep{DavidISW}:
    \begin{equation}\label{cl}
        C_l=\frac{2}{\pi}\int dk k^2 P\left(k\right)\left[W_l\left(k\right)\right]^2 \,,
    \end{equation}
    with 
    \begin{equation}\label{Window}
        W_l\left(k\right)=\int dz n\left(z\right)b\left(z\right)D\left(z\right)j_l\left[k r\left(z\right)\right] \,.
    \end{equation}
    where $C_l$ is the angular power spectrum coefficient of multipole $l$, $W_l\left(k\right)$ is the window function of wavenumber $k$, $b\left(z\right)$ is the bias of the tracers at redshift $z$, $D\left(z\right)$ is the growth factor, $n\left(z\right)$ is the number density, $j_l$ is the Bessel function of order l, and $r\left(z\right)$ is the co-moving distance to redshift $z$. The theoretical power spectrum $P\left(k\right)$ was derived using \ac{CAMB}.
The \ac{ACF} $w\left(\theta\right)$ was then calculated from this theoretical \ac{APS} prediction using equation \ref{legendre}, through a Legendre transform:
\begin{equation}\label{legendre}
    w\left(\theta\right)=\sum c_lP_l\left(\cos\left(\theta\right)\right) \,,
\end{equation}
with 
\begin{equation}
    c_l=\frac{\left(2l+1\right)C_l}{4\pi} \,.
\end{equation}
    \subsection{\ac{ACF}}\label{ACF theory}
    In essence, the \ac{ACF} measures the clustering at each angular scale present in the data set, when compared to that expected if the tracer positions were random. Formally, following the form introduced by \citet{1980lssu.book.....P};
    \begin{equation}
    \delta P=N^2\delta \Omega_1 \delta \Omega_2\left(1+w\left(\theta\right)\right)\,,.
    \end{equation}
    where $w\left(\theta\right)$ is the \ac{ACF}, pertaining to the probability $\delta P$ of two sources located in both solid angles $\delta \Omega_1$ and $\delta \Omega_2$ separated by an angle $\theta$. Thus, the \ac{ACF} formally relates to the probability overdensity, indicative of the source clustering: in the trivial case where $w$ is uniformly zero, one can see that this reduces to the source density multiplied by the area.

    Various estimators for the \ac{ACF} have been proposed, among the most intuitive being that derived from Monte Carlo integration \citep{1980lssu.book.....P,Landy}:
    \begin{equation}
        w\left(\theta)\right)=DD/RR-1 \,.
    \end{equation}
    In this work, that of \citet{Landy} is used, due to the lower variance \citep{Blake2002,Landy}:
    \begin{equation}
        w\left(\theta\right)=\frac{DD}{RR}-\frac{2DR}{RR}+1 \,,
    \end{equation}
    where \textit{DD} is the number of galaxy pairs in the sample at an angular distance \begin{math}
    \theta\Longrightarrow\theta+\delta\theta\,
    \end{math}, 
    \textit{RR} is the number in a random sample, and \textit{DR} is the number of pairs found between the catalogue and random sources.

\section{Methodology}\label{methodology}
    \subsection{Data}\label{data}
    The \ac{GLEAM-X} survey was performed with the \acl{MWA} \citep[MWA; ][]{Tingay2013,Wayth2018}, surveying the sky south of $+30$ degrees declination. The survey has observed the frequency range of 72-231 MHz, and the final data will have a resolution varying from $2'$ to $45''$, with a resolution of approximately $45''$ in the 170--231 MHz image, used for source-finding. The first data release from the GLEAM-X survey was produced in 2022, covering 4\,h$\leq$RA$\leq$13\,h, and Declination range $-32.7^\circ$ to $-20.7^\circ$ \citep{GLEAMX}. Data release 2 will cover 20\,h40\,m$\leq$RA$\leq$6\,h40\,m, -90$\leq$Dec$\leq$+30 \citep{2024arXiv240606921R}.

\begin{figure}[ht]
    \centering
    \includegraphics[width=0.5\textwidth]{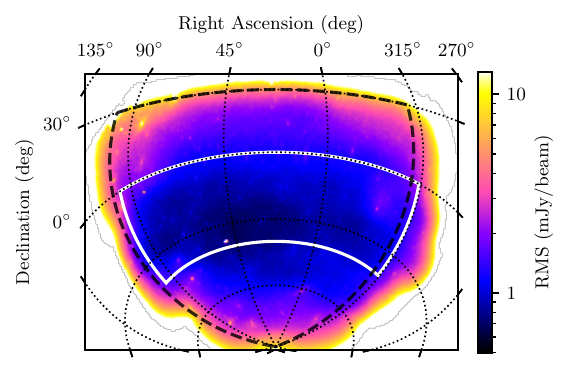}
    \caption{A view of the \ac{GXDS} region, highlighted in the white box, and the rms values. The \ac{GLEAM-X} DR2 region is shown in the black dashed line.}
    \label{RMS}
\end{figure}

    The \ac{ACF} was calculated from a subset of the \ac{GLEAM-X} data. The usable data was reduced by highly sporadic noise properties near the edges of the image, leading to variations in source density. As discussed by \citet{Blake2002}, due to the dependence of the galaxy data pairings on the local source density, i.e on the amount of clustering, differing with that of the randoms, which is dependent on the global density, the \ac{ACF} can be artificially increased. As such, a mask was applied to the data, to excise these regions from the image, resulting in sections between RA $21^h 4^m$ to $6^h 24^m$, and Dec $-40^\circ$ to $0^\circ$ being used. This is shown in figure~\ref{RMS}, and shall hereby be referred to as \ac{GXDS}. This region contains 362944 sources, found in the source-finding image by the \textsc{AEGEAN} source finding package \citep{2018PASA...35...11H} as part of \ac{GLEAM-X} processing, and has a mean rms value of $1 \mathrm{mJy\ beam^{-1}}$. \textsc{AEGEAN} uses a novel source finding technique, namely priorised fitting, whereby the location of a source is determined in the source finding image, and then used at lower frequencies to inform where the source should be \citep{GLEAMX}.

The completeness of GLEAM-X DR2 is discussed by \cite{2024arXiv240606921R}.
The survey is approximately 90\% complete in this region for a flux density level of 10~mJy. As such, a flux density cut of 10~mJy was applied to both the randoms and data, the cut being performed at the source-finding image central frequency of 200~MHz.

    \subsection{Source counts}
\begin{figure}[ht]
        \centering
        \includegraphics[width=0.5\textwidth]{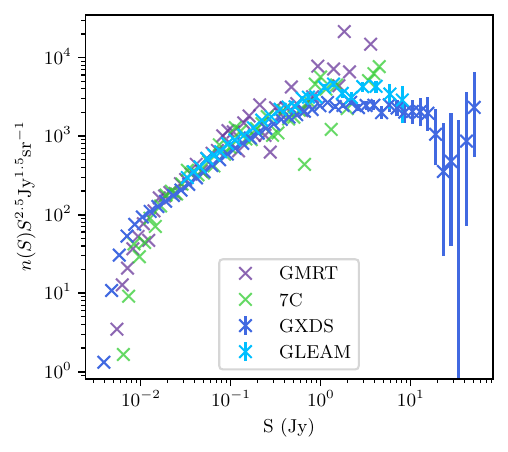}
        \caption{The normalised Euclidean source counts of the \ac{GXDS} region, compared to other surveys, namely the the 151-MHz 7C survey \citep{Hales}, 200-MHz counts from the \ac{GLEAM} survey \citep{Franzen}, and source counts from \ac{GMRT} observations of the Bo\"{o}tes field \cite{2011A&A...535A..38I}. The source counts have not been corrected for frequency scaling.}
        \label{counts}
    \end{figure}

    The \ac{GLEAM-X} \ac{GXDS} differential source counts are shown in Figure \ref{counts}. This uses the \ac{GXDS} region defined in section \ref{data}, with 200610 sources after flux density cutting. One can see good agreement with other surveys, namely that of the 151-MHz 7C survey \citep{Hales}, 200-MHz counts from the \ac{GLEAM} survey \citep{Franzen}, and source counts from \ac{GMRT} observations of the Bo\"{o}tes field \citep{2011A&A...535A..38I}. It is also evident from the counts that the survey drops considerably in completeness below 10\,mJy. This is reflected in a flux density cut applied to the data (see section \ref{data}) before measurements of the source counts and \ac{ACF} were made. The source counts are tabulated in appendix \ref{source_counts}
    \subsection{Angular Correlation Function} \label{ACF methodology}
    Following this, random sources were generated, to compare the clustering to that present in the data. As well as a random position, these sources were assigned a flux density, this value being derived from the $n(s)$ distribution for the GLEAM-X data. In an effort to replicate the effect of noise on surface density in the remaining image, the random sources were retained if the flux density, which when a normally distributed noise component was added, was 5$\sigma$ above the noise, as per the process used in GLEAM-X source finding. A further flux density cut of 10\,mJy was made, to ensure the uniformity of the source density, as discussed in section \ref{data}. 
    
    The \ac{ACF} of the resulting data was then calculated. The measurement of the pairs at a given angular separation was done using the \textsc{TreeCorr} \citep{TreeCorr} package, for both the GLEAM-X galaxies and the generated random catalogues.
    
    \subsection{Window function, redshift distribution and n(z)}\label{window etc}
\begin{figure}
        \centering
        \includegraphics[width=0.5\textwidth]{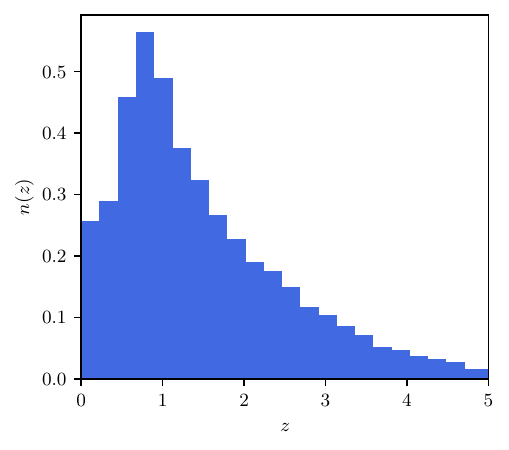}
        \caption{The \ac{SKADS} approximated distribution of GLEAM-X redshifts, as discussed in section~\ref{window etc}.}
        \label{nz}
    \end{figure}

    In translating the 3-dimensional clustering prediction to the observed \ac{ACF}, one must assume some information about how the sources are clustered in redshift space. To make an informed prediction, the \acl{SKADS} \citep[SKADS;][]{2008MNRAS.388.1335W} simulations were used: these simulations were conducted to provide a dataset resembling that which may eventuate from the \acl{SKA} \citep[SKA;][]{SKA}, so astronomers could test science cases with a realistic dataset. A flux density cut of 10\,mJy was applied to the \ac{SKADS} estimate, scaled to the \ac{SKADS} frequency of 1.4GHz, and the redshift distribution of the resulting sources was chosen to approximate that in \ac{GXDS}. The $n\left(z\right)$ distribution is shown in Figure~\ref{nz}.
    
    \subsection{Theoretical estimation and bias fitting.}\label{theoretical estimation}
    As mentioned above, the observed angular power spectrum and \ac{ACF} are related to the fundamental parameters of our universe. As an intent of our work is to see if the measurements made using the GLEAM-X data are consistent with the currently accepted $\Lambda$CDM cosmological values, summarised in table~\ref{cosmo_params}.
    \begin{table*}[]
        \centering

        \begin{tabular}{c|c|c|c|c|c|c}
            Parameter & $n_s$ & $H_0$ & $\Omega_{\mathrm{CDM}}$ & $\Omega_{b}$ & $A_s$ & $\tau$  \\
            \hline
            Value & 0.965 & 67.37 & 0.264 & 0.049 & $2.4\times 10^{-9}$ & 0.054
        \end{tabular}

        \caption{The cosmological parameters used in this analysis}
        \label{cosmo_params}
    \end{table*}
    We set the default cosmology for theoretical predictions to these values. Indeed, degeneracy in the computations involving bias and $n(z)$, as demonstrated in equation~\ref{Window}, mean that we would be unable to measure the values exclusively from the GLEAM-X data without assuming the value of arbitrarily chosen parameters.
    The chosen values, from \cite{Plank2020cosmo}, were used to generate a theoretical power spectrum and correlation function, thereby allowing theoretical comparison with the observed data.
    \\

    Attempts were made in this work to derive the bias (see section \ref{Bias}) of the source population, primarily \ac{AGN} in \ac{GLEAM-X} \citep{2019PASA...36....4F}, and check consistency with various theoretically motivated and parametric models \citep{LoTSSW,DavidISW,Blake_2004_C_l_NVSS}. To do this, the \ac{ACF} and angular power spectrum, assuming best-fitting \textit{Planck} cosmology, were theoretically estimated with multiple bias models, and the model that best fit the observational data identified. The models used are listed in Table~\ref{bias_table}.
    \subsection{Covariance estimation}\label{covariance}
    In estimating the covariance of the angular power spectrum, including both that due to cosmic variance and statistical variance, three techniques were used: two internal and one external. The covariance matrix was calculated analytically from the data, using the Jackknife methodology of \citet{Norberg_2009}, and externally through the variance of generated mock galaxy catalogues. This was necessary due to the effects of cosmic variance not being measured by internal covariance estimators, such as that calculated analytically \citep{DavidISW,2001ApJ...546....2E}.
        \subsubsection{Jackknife estimation}\label{Jackknife covariance}
    The first method of estimating the covariance matrix was by 'jackknife' estimation \citep{Norberg_2009}. 50 subsamples of the healpix cells to which the sky distribution was binned were drawn, each sample consisting of non adjacent healpix cells exclusive to that subsample. The \ac{ACF} of the sky excluding each sub-sample was then calculated for all subsamples. The overall covariance matrix was then estimated as:
    \begin{equation}
        \hat{C}_{\theta\theta'}=\frac{N_{sub}-1}{N_{sub}}\sum_{i=1}^{N_{sub}}\left(\left\{W_{\theta}\right\}_i-\Bar{W}_\theta\right)^{T}\left(\left\{W_{\theta'}\right\}_i-\Bar{W}_{\theta'}\right) \,,
    \end{equation}
    with $W_\theta$ the correlation function at angle $\theta$ , and: 
    \begin{equation}
    \Bar{W}_\theta=\frac{1}{N_{sub}}\sum_{i=1}^{N_{mock}}\left\{W_{\theta}\right\}_i \,.
    \end{equation}
    \subsubsection{Analytic covariance}\label{analytic covariance}
    The covariance matrix was also estimated analytically following \cite{2011MNRAS.414..329C}. That is, the analytic covariance between two angles $\theta$ and $\theta'$ is represented by:
    \begin{equation}
        Cov_{\theta\theta'}=\frac{2}{f_{sky}}\sum_{l\geq0}\frac{2l+1}{\left(4\pi\right)^2}P_l\left(\cos{\theta}\right)P_l\left(\cos{\theta'}\right)\left(C_l+\frac{1}{\hat{n}}\right)^2 \,,
    \end{equation}
    with $C_l$ estimated as per equation \ref{cl}, $\hat{n}$ the source density per steradian, and $f_{sky}$ the fraction of sky covered by the survey.
    \subsubsection{Covariance of mock samples}\label{sample covariance}
    In order to quantify the effect of cosmic variance, and validate the covariance matrix, the covariance was estimated from mock galaxy catalogues, generated from the same underlying cosmology as that measured by \citet{Planck_Cl}, and consistent with the GLEAM-X angular power spectrum and \ac{ACF}s measured in this work. The sample covariance of the mock realisations is calculated as \citep{DavidISW}:
    \begin{equation}
        \hat{C}_{\theta\theta'}=\frac{1}{N_{mock}-1}\sum_{i=1}^{N_{mock}}\left(\left\{W_{\theta}\right\}_i-\Bar{W}_\theta\right)^{T}\left(\left\{W_{\theta'}\right\}_i-\Bar{W}_{\theta'}\right) \,.
    \end{equation}
    In generating these mock samples, the \acl{GLASS} \citep[GLASS;][]{GLASS} software package was used to generate mock catalogues. \ac{GLASS} builds light cones in an iterative manner, given an assumed underlying cosmology, with values in this paper set to those detailed by \cite{Plank2020cosmo}, referenced in Table \ref{cosmo_params}. GLASS uses a hybrid of statistical and physical models, log-normal realisations of the gaussian field, and other relations to produce accurate realisations in reasonable timeframes. We used 500 mocks in our analysis, generated with a maximum multipole $l$ of 24564.
The covariance matrices are discussed in section \ref{covariance results}.

\section{Results}\label{Resukts}
    
\subsection{\ac{ACF}}\label{ACF results}
\begin{figure}[ht]
        \centering
        \includegraphics[width=0.5\textwidth]{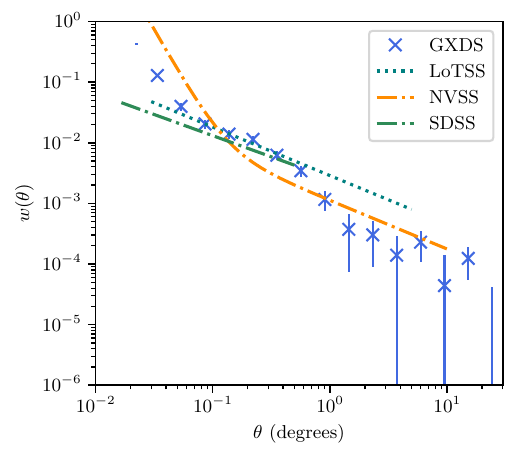}
                \caption{The \ac{ACF} computed with various surveys. The crosses show that computed in this work with the \ac{GXDS} survey data, whilst those from the work of \cite{Blake2002} using the \acl{NVSS}\citep[NVSS;][]{1998AJ....115.1693C}, and more recently that of \cite{2024MNRAS.527.6540H} using \acl{LoTSS} \citep[LoTSS; ][]{LoTSS}. Also shown is a result by \cite{Connolly2002} using the \acl{SDSS} \citep[SDSS; ][]{YorkSDSS}. Note that the three fits displayed make use of Limber's approximation \citep{1953ApJ...117..134L}, and model the \ac{ACF} as a power law.}
        \label{plain_ACF}
    \end{figure}

    The \ac{ACF} derived from the \ac{GLEAM-X} is displayed in Figure~\ref{plain_ACF} with errors derived from the Jackknife covariance matrix (see section \ref{Jackknife covariance}). The form of the \ac{ACF} resembles that of prior results; in particular, it displays the double power law morphology of the ACF produced by \cite{Blake2002}: at a similar angular scale to \cite{Blake2002} we see an increase in the slope of the correlation function. In \cite{Blake2002} this is attributed to multiple-component sources: we attribute this to similar features, as discussed in section~\ref{multiple_component_sources}. 

    It is important to note that the fitting of the \ac{ACF} as a power law, as in works by \cite{Blake2002} and \cite{2024MNRAS.527.6540H} relies on Limber's approximation \cite{1953ApJ...117..134L}, which is only applicable for small angular scales. Below the scale of 1$^\circ$ we see excellent agreement with the \ac{ACF} presented by \cite{2024MNRAS.527.6540H}.
    
    One also sees similar cosmological slope (i.e the slope of the power-law component not due to multiple component sources, seen at angular scales of about $~0.1$ degrees) to \cite{2024MNRAS.527.6540H} and \cite{2002ApJ...579...42C}, but a difference in amplitude, and an overall curvature at higher angular scales. The amplitude of the \ac{ACF} is dependent on flux density, luminosity and source type \citep{2024MNRAS.527.6540H}, and as this differs for the different surveys, we would not expect this to match exactly. This is further discussed in section \ref{results_bias_fits}.
\subsection{Systematic effects}
    As per prior studies such as that by \cite{2024MNRAS.527.6540H}, systematic effects in the data can give the measure of spurious clustering in the resulting \ac{ACF}. The GLEAM-X data used in this work has the potential to be affected by three effects, namely flux density scaling, calibration errors and ionospheric effects that could smear out the sources, and primary beam dependent effects. These are addressed in turn.
    \subsubsection{Flux density scaling}\label{flux density scaling}
    The flux density scale for \ac{GLEAM-X} was computed using a model from the precursor survey \ac{GLEAM}. \ac{GLEAM} in turn has a flux density scale derived from the \ac{NVSS}, \acl{SUMSS} \citep[SUMSS; ][]{1999AJ....117.1578B}, and the \acl{VLSSr} \citep[VLSSr; ][]{2014MNRAS.440..327L}, averaged over scales of $\sim10,000$sq.deg. . As detailed by \cite{2017MNRAS.464.1146H}, this leads to an internal flux density scale that is free of Declination-dependent effects down to a Declination of $-72^\circ$, which was also tested with independently-calibrated VLA observations of various well-known calibrator sources. Both GLEAM and GLEAM-X processing are performed in drift scans with very large primary beam overlaps, leading to minimal variation in the flux density scale with Right Ascension. The sub-band channels are processed independently using the same methods, and the final radio source spectra derived from these measurements are consistent with those expected from other studies i.e. a median spectral index of $\alpha=-0.83$, where flux density $S\propto\nu^\alpha$, for bright, isolated AGN \citep{2017MNRAS.464.1146H,GLEAMX}. We therefore have no cause to suspect any issues with the flux density scale that could cause any systematics in the measured clustering.
    \subsubsection{Smearing}
    A potential issue in the dataset, that has affected other results \citep{2024MNRAS.527.6540H}, is that of smearing, namely sources being altered in dimension such that the peak flux density is reduced, and the source subsequently not detected. To gauge whether this is an issue for GLEAM-X, we consulted the rms maps for the region considered, and reviewed the procedures that went into making the GLEAM-X mosaics, as detailed by \cite{2024arXiv240606921R}). Briefly, a catalogue of sparse, unresolved, and high signal-to-noise sources from \ac{NVSS} and \ac{SUMSS} is constructed, and the ratio of the integrated to peak flux density for those sources in GLEAM-X is calculated. A mean greater than or equal to $1.1$ or a standard deviation greater than or equal to $0.125$ of $S_{int}/S_{peak}$ saw the observation flagged and not included in the final mosaics. Furthermore, in constructing the \ac{GLEAM-X} mosaics, ionospheric corrections were applied to the positions of the sources, with the catalogue of \ac{NVSS} and \ac{SUMSS} sources used to derive a model of position shifts \citep{GLEAMX}. These were applied to every image before mosaicing. Only unresolved, single component sources were used in these corrections. Finally, a 5 sigma applied to the data and randoms should ensure that any smearing remains minimal.
    \subsubsection{Primary beam effects and other  effects}
    To further see if factors such as flux density scaling and position dependent affects contribute to the uncertainty of the \ac{ACF}, the area used in this study was divided into 14 regions, each square in RA and Dec and 20 degrees on a side, and the \ac{ACF} calculated from each region. This is presented in figure \ref{region_acf}, with the \ac{ACF}'s presented separately in appendix 2, figure \ref{ACFDF_subset}.
    \begin{figure}
        \centering
        \includegraphics[width=0.5\textwidth]{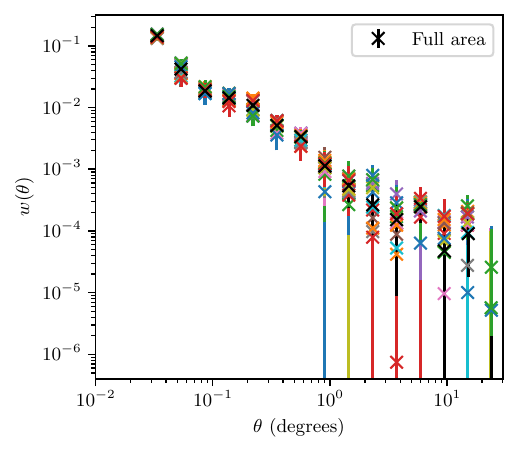}
        \caption{The \ac{ACF} calculated for each sub-region of the \ac{GXDS} dataset.}
        \label{region_acf}
    \end{figure}
    Particularly on the small scales, one can see that the calculated \ac{ACF} is similar for all regions, diverging somewhat at scales over 10 degrees. However, the variation is within the errors from the full area. There does not seem to be an observed trend with either RA or Dec.
    
    The variation about the \ac{ACF} calculated using the full \ac{GXDS} region is significant for scales over approximately $1^\circ$. The jackknife covariance matrix was calculated using 50~patches, and as such the sky area covered in each patch will be smaller, however variation in source density over different sky patches could still significantly contribute to the jackknife covariance at larger scales. 
\subsection{Bias fits}\label{results_bias_fits}
\begin{figure}[ht]
        \centering
        \includegraphics[width=0.5\textwidth]{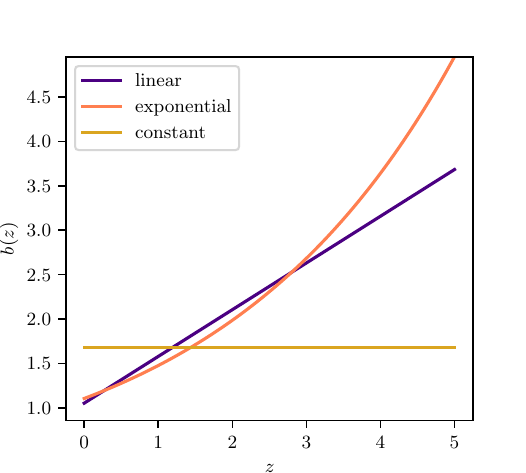}
        \caption{The three empirically-fit bias distributions of Table \ref{bias_table}, used in computing the theoretical bias fits shown in Figure~\ref{bias_fits_ACF}.}
        \label{bias_fits}
    \end{figure}

\begin{figure}[ht]
        \centering
        \includegraphics[width=0.5\textwidth]{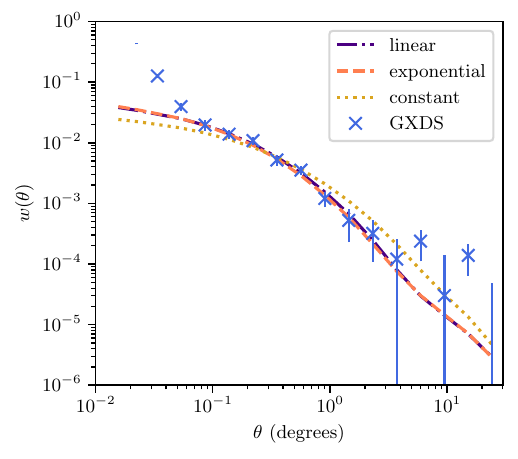}
        \caption{The theoretical \ac{ACF} derived from various bias fits, with the observational \ac{ACF} for comparison, as discussed in section \ref{results_bias_fits}. The plotted data is identical to that in Figure \ref{plain_ACF}.}
        \label{bias_fits_ACF}
    \end{figure}

    As discussed in section~\ref{Bias}, when using discrete tracers of the underlying matter distribution, one needs to be mindful of their bias. An attempt was made in this work to derive the bias of the \ac{AGN} observed by the \ac{GLEAM-X} survey. Four different bias models were fitted. The first was the bias model discussed by \cite{2015ApJ...812...85N} and used in \cite{LoTSSW}:
    \begin{equation}
        b\left(z\right)=1.6+0.85z+0.33z^2 \,,
    \end{equation}
    and three parametric models, namely:
    \begin{equation}
        b\left(z\right)=az+b_0 \,,
    \end{equation}
    \begin{equation}
        b\left(z\right)=ae^{b_0\times z} \,,
    \end{equation}
    \begin{equation}
        b\left(z\right)=b_0 \,.
    \end{equation}
    The best-fitting values of the parametric models are displayed in Table~\ref{bias_table}, and plotted in Figure~\ref{bias_fits}. The corresponding theoretical \ac{ACF}'s are plotted in Figure~\ref{bias_fits_ACF}. The bias models were fitted excluding the first 4 data points, covering scales to $~0.1^\circ$, for the uptick in the gradient of the \ac{ACF} evident in these points can be attributed to multiple component sources, as discussed in section~\ref{multiple_component_sources}.
    \begin{table}[ht]
        \centering
        \begin{tabular}{c|c|c|c}
            model & $b_0$ & $\frac{db}{dz}$ or $\frac{d\ln{b}}{dz}$ & reduced $\chi^2$ \\
            \hline
           linear & $1.05 \pm 0.25$ & $0.526 \pm 0.38$ & $0.764$ \\
           exponential  &  $0.11 \pm 0.15$ & $0.316 \pm 0.07$  & $0.827$ \\
           constant & $1.58 \pm 0.03$ & -- & $2.66$ \\
        \end{tabular}
        \caption{The best fitting bias fits for the \ac{GLEAM-X} \ac{ACF}, plotted in Figure~\ref{bias_fits}}
        \label{bias_table}
    \end{table}

    We find that the model of \cite{2015ApJ...812...85N} is not consistent with our \ac{ACF}. Of course, given the degeneracy for the $n\left(z\right)$ and the $b\left(z\right)$ distributions in computing the \ac{ACF}, and the fact that the redshift distribution of the \ac{GLEAM-X} galaxies was approximated from \ac{SKADS}, this may simply be representative of our redshift distribution being uncertain.
    \begin{figure}[ht]
        \centering
        \includegraphics[width=0.5\textwidth]{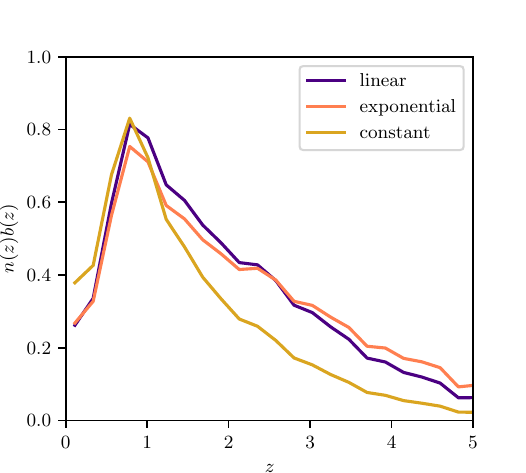}
        \caption{$n\left(z\right)b\left(z\right)$ for the various best fit bias fits described in Section \ref{results_bias_fits}. The three bias fits, namely linear, exponential and constant, are of the form $b\left(z\right)=a\times z+b_0$, $b\left(z\right)=a\times e^{b_0\times z}$ and   $b\left(z\right)=b_0$ respectively, with $n\left(z\right)$ derived from \ac{SKADS}.}
        \label{nzbz}
    \end{figure}

\subsection{Multiple component sources}\label{multiple_component_sources}
When viewing the \ac{ACF} derived from the \ac{GLEAM-X} survey, one can see an increase in the gradient at angular separations lower than $~0.1$ degrees. This phenomenon is seen in prior studies, such as that by \cite{Blake2002}, and is representative of sources with multiple components being included in the analysis, spuriously enhancing the clustering. Given \ac{GLEAM-X} has the same resolution as the \ac{NVSS}, the survey used in \cite{Blake2002}, it is reassuring we see a similar feature.

When compared to \cite{Blake2002}, one can see the gradient of the \ac{ACF} caused by the multiple component sources is lower in our study. However, when we re-compute the \ac{ACF} after a flux density cut of 50mJy, consistent with the \cite{Blake2002} study with an assumed spectral index of $-0.83$, we recover the same \ac{ACF} as \cite{Blake2002}. With the extra sensitivity of the \ac{GLEAM-X} survey, we are detecting fainter sources which are less resolved, thus decreasing the proportion of sources that have multiple components, and reducing the gradient of our \ac{ACF} at scales below $0.1 ^\circ$.
\subsection{Covariance measures}\label{covariance results}
\begin{figure*}[ht!]
        \centering
        \includegraphics[width=\textwidth]{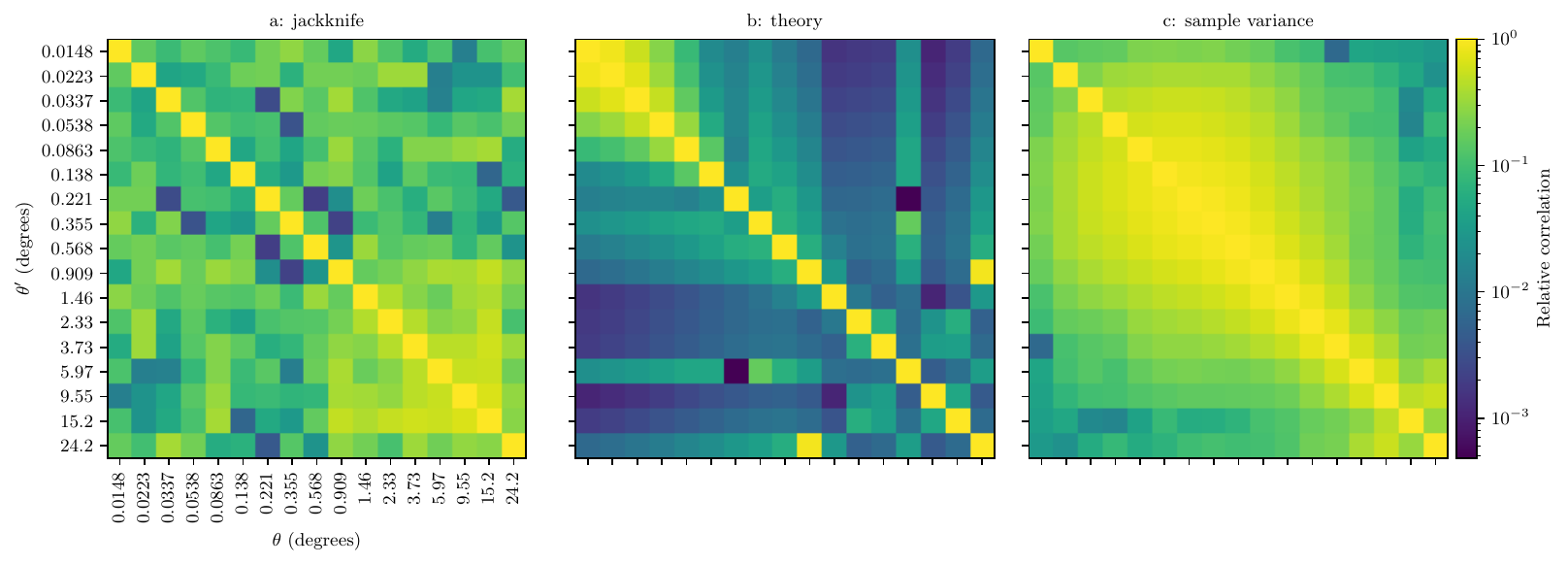}
        \caption[width=\textwidth]{\label{triple_corr}The three correlation matrices for the analysis. That computed by the the jackknife methodology in section \ref{Jackknife covariance} is on the left, the centre figure concerns the theoretical methodology discussed in section \ref{theoretical estimation}, and the right figure depicts that computed using the sample variance methodology in Section~\ref{sample covariance}.}
    \end{figure*}

 The correlation matrices for the analysis are shown in Figure \ref{triple_corr}. As prefaced in section \ref{covariance}, this allows the estimation of the contributions of cosmic and statistical variance to the errors of the measured ACF.
    
    In viewing the jackknife correlation matrix estimate in Figure \ref{triple_corr}, one is struck by two evident features. The first is the expected higher amplitude on the diagonal, expected as the variance of bins will be higher than the covariance of one bin with a different bin. The second feature is the large amount of correlation at the larger angular scales, from approximately $1^\circ$. This can be attributed to two effects, both limiting the amount of independent samples at higher angular scales. The first is the limited sky coverage, imposed by our survey mask, and the second is the source density, resulting in their being less pairs to sample from. The first of these is equivalent to the approximation, present in multiple prior studies \citep{2001ApJ...546....2E,1994ApJ...421L...5S} that the error in the angular power spectrum scales with $\frac{1}{f_{sky}}$.
    \\
    Interestingly, the theoretical correlation matrix in figure \ref{triple_corr}b, does not show the same large correlation at high angular scales, however at small scales the theoretical correlation matrix shows an increase in the relative cross-correlation of the measurements. The lack of the large-scale correlations is not surprising, as the theoretical matrix computations do not take into account the shape of the survey mask. The matrix has far lower off-diagonal components than either the jackknife or sample variance matrix, differing by several orders of magnitude, indicating that these off-diagonal elements are greatly affected by survey related matters such as the survey mask, as well as cosmic variance.
    \\
    \\
    The final of the three correlation matrices, that relating to the sample variance methodology, shows a mixture of two features and is displayed in figure \ref{triple_corr}c. At small scales, we have the effects of limited sky density, and a high-resolution simulation, resulting in many empty cells in the pixelisation scheme, and in correlation at small scales. On the large-scales, we have a similar feature to that in the jackknife correlation matrix, namely the survey mask resulting in larger scales being correlated.
\section{Conclusion}\label{conclusion}
\begin{itemize}
    \item We use a subset of the \ac{GLEAM-X} survey, corresponding to RA $21^h 4^m$ to $6^h 24^m$, and Dec $-40^\circ$ to $0^\circ$ and containing 200610 sources after flux cutting, to measure the \ac{ACF} clustering of the present galaxies.
    \item  A good handling of error estimation allows us to estimate the covariance of the \ac{ACF} measurements. In computing the covariance by theoretical, jackknife and sample variance means, we find that the results are similar, however the theoretical prediction fails to represent effects of either cosmic variance or the peculiarity of the dataset.
    \item We find the cosmological properties represented by the measured \ac{ACF} to be consistent with the $\Lambda$CDM model of cosmology, assuming that the values of the cosmological parameters are accurately measured by \citep{Plank2020cosmo}.
    \item We fit different models of the bias, the parameter that relates the amplitude of the measured ACF for the AGN tracers we observe with that of the underlying matter distribution. We find that the best fit comes from a bias that evolves, either linearly or exponentially, with redshift, with a bias constant in redshift space being a poorer fit, consistent with prior studies such as \cite{2024MNRAS.527.6540H,2024A&A...681A.105N}.
    \item The cosmological utility of \ac{GLEAM-X} shall only be improved as more sky coverage is attained. This shall allow not only smaller covariance between angular scales, but the measurement of the \ac{APS}, as well as cross correlation with other data, such as the \ac{CMB}.
    
\end{itemize}
\begin{acknowledgements}

\end{acknowledgements}
BV acknowledges a Doctoral
Scholarship and an Australian Government Research Training Programme scholarship administered through Curtin University of
Western Australia. 
NHW is supported by an Australian Research Council Future Fellowship (project number FT190100231) funded by the Australian Government.

This scientific work uses data obtained from Inyarrimanha Ilgari Bundara / the Murchison Radio-astronomy Observatory. We acknowledge the Wajarri Yamaji People as the Traditional Owners and native title holders of the Observatory site. Establishment of CSIRO's Murchison Radio-astronomy Observatory is an initiative of the Australian Government, with support from the Government of Western Australia and the Science and Industry Endowment Fund. Support for the operation of the MWA is provided by the Australian Government (NCRIS), under a contract to Curtin University administered by Astronomy Australia Limited. This work was supported by resources provided by the Pawsey Supercomputing Research Centre with funding from the Australian Government and the Government of Western Australia.

This work makes use of the Numpy \citep{Numpy}, Matplotlib \citep{Matplotlib}, Astropy \citep{astropy:2013,astropy:2018,astropy:2022} and Pandas \citep{pandas} python packages.
\begin{appendix}\label{source_counts}
\section{Source counts}
\begin{table*}
\caption{The tabulated source counts at 200\,MHz from the \ac{GXDS} region, displayed in figure \ref{counts}. The highest flux density counts are incomplete due to small number statistics.}
\begin{tabular}{rrr}
\hline
 bin centre (Jy) &  bin width (Jy) &  Euclidian normalised counts $n(S)S^{2.5} (\mathrm{Jy}^{1.5} $sr$^{-1})$ \\
\hline
   0.003911 &   0.000770 &                     1.227996 \\
   0.004765 &   0.000938 &                     9.626755 \\
   0.005806 &   0.001143 &                    27.380452 \\
   0.007073 &   0.001392 &                    49.540736 \\
   0.008618 &   0.001696 &                    71.676393 \\
   0.010499 &   0.002067 &                    91.707164 \\
   0.012791 &   0.002518 &                   109.111880 \\
   0.015584 &   0.003068 &                   126.667409 \\
   0.018987 &   0.003737 &                   147.868935 \\
   0.023132 &   0.004553 &                   175.278653 \\
   0.028183 &   0.005548 &                   205.631163 \\
   0.034336 &   0.006759 &                   243.075005 \\
   0.041832 &   0.008234 &                   293.771879 \\
   0.050966 &   0.010032 &                   351.954891 \\
   0.062093 &   0.012223 &                   423.281635 \\
   0.075650 &   0.014891 &                   500.900254 \\
   0.092167 &   0.018142 &                   591.303146 \\
   0.112290 &   0.022103 &                   700.481778 \\
   0.136806 &   0.026929 &                   808.723700 \\
   0.166676 &   0.032809 &                   946.078425 \\
   0.203066 &   0.039972 &                  1092.454673 \\
   0.247402 &   0.048699 &                  1254.995441 \\
   0.301417 &   0.059332 &                  1424.931702 \\
   0.367226 &   0.072286 &                  1634.100437 \\
   0.447404 &   0.088068 &                  1749.836152 \\
   0.545086 &   0.107296 &                  1871.024310 \\
   0.664096 &   0.130723 &                  2071.879783 \\
   0.809089 &   0.159264 &                  2181.911126 \\
   0.985738 &   0.194036 &                  2444.108267 \\
   1.200956 &   0.236400 &                  2686.138440 \\
   1.463163 &   0.288014 &                  2372.636427 \\
   1.782618 &   0.350896 &                  2466.532721 \\
   2.171820 &   0.427508 &                  2698.168496 \\
   2.645997 &   0.520846 &                  2305.273390 \\
   3.223702 &   0.634564 &                  2439.695783 \\
   3.927538 &   0.773109 &                  2442.122985 \\
   4.785044 &   0.941903 &                  1924.015791 \\
   5.829770 &   1.147550 &                  2498.141118 \\
   7.102594 &   1.398097 &                  2159.631948 \\
   8.653315 &   1.703346 &                  1936.140129 \\
  10.542608 &   2.075240 &                  1952.748675 \\
  12.844393 &   2.528331 &                  2042.444125 \\
  15.648731 &   3.080345 &                  1961.871014 \\
  19.065345 &   3.752882 &                  1055.306818 \\
  23.227913 &   4.572255 &                   354.786474 \\
  28.299302 &   5.570523 &                   477.106526 \\
  34.477936 &   6.786745 &                     0.000000 \\
  42.005562 &   8.268506 &                   862.803624 \\
  51.176707 &  10.073783 &                  2320.546409 \\
\hline
\end{tabular}
\end{table*}
\end{appendix}
\begin{appendix}\label{acfdf_subset}
\section{}
\begin{figure}
    \centering
    \includegraphics[width=0.5\textwidth]{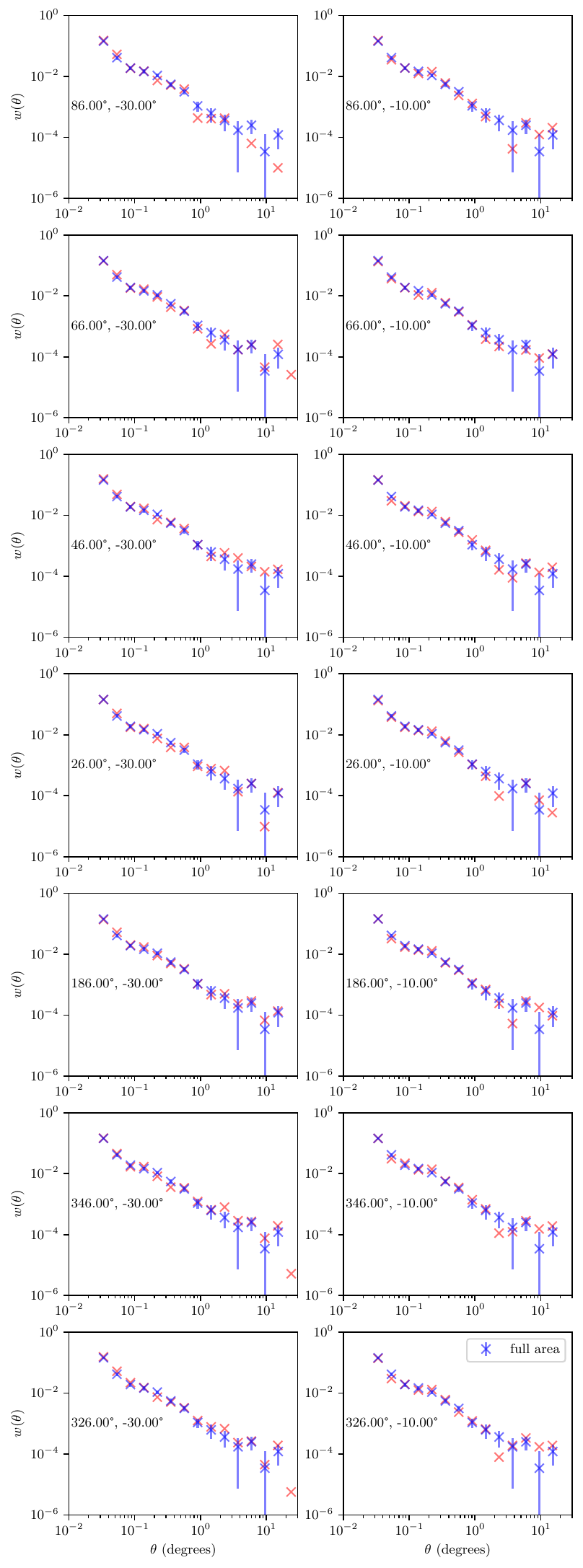}
    \caption{The \ac{ACF} plotted from each subset of the \ac{GXDS} region used. The subsets were 20 degrees on a side, with the centre RA and Dec of each listed on the plot.}
    \label{ACFDF_subset}
\end{figure}
\end{appendix}
\twocolumn
\bibliographystyle{pasa-mnras}
\bibliography{example}

\end{document}